\documentclass[prb,aps,twocolumn,floatfix]{revtex4}
\usepackage{graphicx}% Include figure files
  \usepackage{amsmath}
  \usepackage{subfigure}
  \usepackage{braket}
\usepackage{tikz}
\usetikzlibrary{arrows} 
\usepackage{amssymb}
\usepackage{amsfonts}
\usepackage{bm}
\begin{document}

\newcommand {\be}{\begin{equation}}
\newcommand {\ee}{\end{equation}}
\newcommand {\C}{\mathbb C} % Complex
\newcommand {\Z}{\mathbb Z} % Integers
\newcommand {\R}{\mathbb R} % Real
\newcommand {\I}{\mathbb I} % Real
\newcommand {\F}{\mathfrak F}
\newcommand {\Y}{\mathfrak Y} 
\newcommand{\beq}{\begin{eqnarray}}
\newcommand{\eeq}{\end{eqnarray}}
\newcommand{\bk}{{\bf k}}

\date{\today}
\title{Spin stiffness calculation in anisotropic XY model with Ring exchange interaction.}
\author{Solomon Akaraka Owerre}
%\email{solomon.akaraka.owerre@umontreal.ca}
\affiliation{$^1$Groupe de physique des particules, D\'epartement de physique,
Universit\'e de Montr\'eal,
C.P. 6128, succ. centre-ville, Montr\'eal, 
Qu\'ebec, Canada, H3C 3J7 }

\begin{abstract}
We present the spin wave theory of $XY$ model with anisotropic nearest neighbour (NN) interactions $J(J^{\prime})$ along the $x(y)$ directions, next nearest (NNN) neighbour interaction $J_D$ and the ring exchange interaction $K$ on the square lattice. We calculate the thermodynamic quantities: Zero temperature spin stiffness, internal energy, specific heat and the magnetization. Using the diagonalized Hamiltonian, we show that no soft modes develop when $\eta + \lambda >0 $, where $\eta = J^{\prime}/J$ and $\lambda = K/J$. We further show  that anisotropy ($\eta =2$) decreases the spin stiffness by $5.7\%$ of its isotropic ($\eta =1$) maximum value for some values of $\lambda$ and $\delta = J_D/J$. A similar reduction shows up in the magnetization. The plot of the stiffness against $\eta (\lambda)$ reaches a maximum at $\eta=0(\lambda =0)$ for specific values of $\delta$ and decreases rapidly as it approaches  $\eta + \lambda + 1=0$. In general, the supersolid phase transition suggested earlier\cite{H} will occur in the regime $\eta + \lambda +1 <0$.

\end{abstract}
\maketitle

\section{Introduction} The quantum $XY$ model has been a subject of interest for many decades. It has been intensively studied both numerically and theoretically \cite{A, B, F, H, K, L, Z2}. This model is of considerable importance for the understanding of quantum phase transitions and thermodynamic properties at low temperatures. Most theoretical studies of this model is based on spin wave theory method.

This method involves the mapping of the spin operators to bosonic operators such that the spin commutation relation is satisfied. It provides a good approximation for obtaining the low-lying  excitation spectrum of quantum spin systems. Several versions of spin wave theory exit \cite{A1,A2, J}. The most popular one is based on the Holstein-Primakoff representation \cite{J} which was first applied to the study of Heisenberg model by Anderson \cite{D} and further extended to second order by Kubo \cite{E} and Oguchi \cite{I}. 
 
Gomez-Santos and Joannopoulous \cite{A} showed how this method can be applied to the case of isotropic $XY$ model by making a good choice of the quantization axis. Since then numerous applications of spin wave theory have been carried out on the isotropic $XY$ model with different lattice configurations. Most of the results obtained so far are in a good agreement with quantum Monte Carlo simulations(QMC) \cite{C,K, L}. 

The isotropic $XY$ can be supplemented with other non-trivial interactions such as anisotropic nearest neighbour (NN) interaction, next nearest neighbour (NNN) interaction or the ring exchange interaction. These interactions can lead to exotic quantum phases in the system such as superfluid \cite{G,F}, valance-bond-solid (VBS)\cite{Z1} and gapless Mott insulator (GMI)\cite{X} phases. They can as well destroy the phases depending the strength of the interaction.
 
The ring exchange interaction was first introduced to  study the magnetic properties of solid $^{3}\text{He}$  \cite{M}. This exchange interaction, alone or in competition with the pure $XY$ model (NN exchange only) has been studied extensively on the square lattice using different numerical methods \cite{F, Z1, P}. A linear spin wave theory approximation to this model has also been carried on the square lattice \cite{H}. 

It is well known that the ring exchange interaction destroys the superfluid phase of isotropic $XY$ model \cite{H}. However, in this paper we shall consider the $XY$ model with all the interactions mentioned above: Anisotropic NN, NNN  and the ring exchange interactions. This introduces four (or three dimensionless ) parameters into the Hamiltonian in contrast to just one parameter. The format of the paper is as follows: In Sec.II, we present the model Hamiltonian and the classical ground state. In Sec.III, we apply spin wave theory by choosing our quantization axis along the $x$-direction. Diagonalize the Hamiltonian by means of Bogoluibov transformation and obtain the spectrum. In Sec.IV, we plot the spectrum and the magnetization as a function of the parameters involved. We show that the soft mode of the spectrum imposes a constraint on the parameters of the Hamiltonian.
 In Sec.V, we calculate the zero temperature spin stiffness by applying a twist along the $x$-direction and plot it as a function of the parameters . Finally, in Sec.VI we make some concluding remarks.
 
\section{Model}
\begin{figure}[h]
\centering
\includegraphics[scale=0.20]{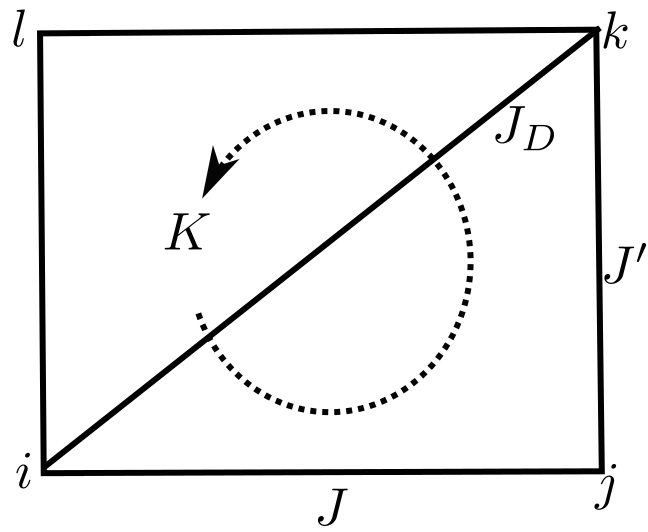}
\caption{(Color online): Actions of the exchange interactions. The hoping strength $J$ acts along the $x$-axis, $J^{\prime}$ acts along the $y$-axis, $J_D$ acts along the diagonal and $K$ is the four spin cyclic ring exchange interaction.} \label{fig3}
\end{figure}

The model Hamiltonian we will consider is of the form
\begin{equation}
\begin{split}
H =&-J\sum_{\left\langle ij \right\rangle} \left(S^{+}_{i}S^{-}_{j} +h.c \right)-J^{\prime}\sum_{\left\langle jk \right\rangle} \left(S^{+}_{j}S^{-}_{k} +h.c \right)\\
&-J_D\sum_{\left[ik \right]}\left(S^{+}_{i}S^{-}_{k} +h.c \right) -K\sum_{\left\langle ijkl  \right\rangle} \left(S_{i}^{+}S_{j}^{-}S_{k}^{+}S_{l}^{-} + h.c\right).
\end{split}  
\label{1}
\end{equation}
where $J$($J^{\prime}$)  is the exchange constants along the $x$($y$) directions respectively, $J_D$ is the exchange constant along the diagonal and $K$ is the ring exchange constant over the square plaquettes (see Fig.\eqref{fig3}).
The spin operators obey the usual commutation relation $\left[S_i^{\alpha},S_j^{\beta}\right]=i\varepsilon_{\alpha \beta \gamma}S_i^{\gamma}\delta_{ij}$, $\alpha,\beta,\gamma =x,y,z$ and $S_{j}^{\pm}= S_{j}^{x} \pm  i S_{j }^{y}$.
A sign problem renders quantum Monte Carlo simulations inapplicable in the regime $K<0$. However, this sign problem cannot be captured by a simple spin wave theory calculation.  In the limit $J =J^{\prime}$ and $J_D =0$, the  Hamiltonian reduces to the one studied in Ref. 10. When $J,J^{\prime}>>J_D,K$, this model undergoes a Kosterlitz-Thouless phase transition \cite{B} at $T_{KT} \approx 0.69$ for 2D model and a superfluid phase for temperatures less than $T_{TK}$ \cite{P}. The regime $J_D=0, K<0$ has been studied in Ref. 23 using variational Monte Carlo (VMC) and density matrix renormalization group (DMRG) method.  

For $J$,$J^{\prime}$,$J_D$,$K >0$, the spins are aligned corresponding to ferromagnetic interactions. All calculations and plots will be done as a function of the dimensionless quantities: $\eta =J^{\prime}/J$, $\delta =J_D/J$, and $ \lambda = K/J$. We will further set $J=1/2$ anywhere else, which corresponds to the parameter value of pure $XY$ model. The classical ground state is found by treating the spins as classical vectors i.e $S_i^{+} = Se^{i\theta}$. For spin-$\frac{1}{2}$ we have
\be
\frac{E_{cl}}{N} = -\frac{J}{2 }\left(1 + \eta + 2\delta + \lambda/4 \right),
\ee
where the coordinate number $z=2$ for NN interactions and $z=4$ for NNN interaction have been used.

% \begin{figure}[h!] 
%\centering
%\begin{tikzpicture}[scale=1.8] 
%\draw[solid, very thick] (0,0)--(2,0) ;
%\draw[solid,very thick] (2,0)--(2,2);
%\draw[solid,very thick] (0,0)--(2,2);
%\draw[solid,very thick] (2,2)--(0,2);
%\draw[solid,very thick] (0,2)--(0,0);
%\draw(-0.1,0) node[]{$i$};
%\draw(1.1,-0.1) node[]{$J_x$};
%\draw(2.1,0) node[]{$j$};
%\draw(2.1, 1) node[]{$J^{\prime}$};
%\draw(1.72,1.6) node[]{$J_D$};
%\draw(2.1,2.002) node[]{$k$};
%\draw(-0.1,2) node[]{$l$};
%\end{tikzpicture}
 
 \section{Spin Wave Theory}
The basic assumption of spin wave theory lies on selecting a classical ground state and determining the fluctuation around it. In other words, one considers quantum fluctuations very close
to an ordered ground state configuration of the system under study.
By choosing the quantization axis along the $x$-direction (instead of the $z$-direction), one can then write the spin operators in terms of the bosonic operators using the famous Holstein-Primakoff transformation\cite{A,J}. Linear spin wave theory corresponds to the transformation
\begin{equation}
\begin{split}
 S_i^x&= \frac{1}{2}-a_i^{\dagger}a_i,\\
 S_i^y &\approx \frac{i}{2}\left(a_i^{\dagger}-a_i\right).
\end{split}
\label{3}
\end{equation}
The spatial Fourier transformation of the bosonic operators is written as \cite{C}
\be
a_i = \frac{1}{\sqrt{N}}\sum_{\bold{k}}e^{i\bold{k}\cdot \bold{R}_i}a_{\bold{k}}, \quad a_i^{\dagger} = \frac{1}{\sqrt{N}}\sum_{\bold{k}}e^{-i\bold{k}\cdot \bold{R}_i}a_{\bold{k}}^{\dagger},
\label{3a}
\ee
where the momentum $\bold{k}$ runs over the first Brillouin zone (BZ) of a square lattice with unit nearest-neighbour distance i.e $\pi< k_x \leq \pi$, $\pi< k_y \leq \pi$. Transforming Eq.\eqref{1}    in terms of $S_{i}^{x}, S_{i}^{y}$  and substituting Eq.\eqref{3} and Eq. \eqref{3a}, the quadratic (non-interacting) part can be grouped in the following way:
\begin{align}\label{eqn3.16}
 \begin{split}
  H_0 = H_{MF} +\sum_{\bold{k}} \left[ A_{\bold{k}}\left(a_{\bold{k}}^{\dagger}a_{\bold{k}} + a_{-\bold{k}}^{\dagger}a_{-\bold{k}} \right) \right.\\
  \left. + B_{\bold{k}}\left(a_{\bold{k}}^{\dagger}a_{-\bold{k}}^{\dagger} + a_{\bold{k}} a_{-\bold{k}} \right) \right].
\end{split}
\end{align}

The Hamiltonian is diagonalized by the Bogoluibov canonical transformation to quasiparticle boson operators $\alpha_{\bold{k}}$ and $\alpha_{\bold{k}}^{\dagger}$ \cite{C,H}:
\be
a_{\bold{k}} = l_{\bold{k}}\alpha_{\bold{k}}-m_{\bold{k}}\alpha_{\bold{-k}}^{\dagger}, \quad a_{\bold{k}}^{\dagger} = l_{\bold{k}}\alpha_{\bold{k}}^{\dagger}-m_{\bold{k}}\alpha_{\bold{-k}},
\label{3b}
\ee
with
\be
l_{\bold{k}} = \sqrt{\frac{A_{\bold{k}} + \varepsilon_{\bold{k}}}{2\varepsilon_{\bold{k}}}}, \quad m_{\bold{k}} = \sqrt{\frac{A_{\bold{k}} - \varepsilon_{\bold{k}}}{2\varepsilon_{\bold{k}}}}.
\label{3c}
\ee
Applying the above transformations, the diagonalized quadratic (non-interacting) part yields
\be
H_0 = H_{MF} + \sum_{\bold{k}}\left(\varepsilon_{\bold{k}}-A_{\bold{k}}\right)+ \sum_{\bold{k}}\varepsilon_{\bold{k}}(\alpha_{\bold{k}}^{\dagger}\alpha_{\bold{k}} + \alpha_{\bold{-k}}^{\dagger}\alpha_{\bold{-k}}).
\label{3d}
\ee
%\omega_{\bold{k}}= \sqrt{A_{\bold{k}}^{2}-B_{\bold{k}}^{2}},\\\label{eqn2} 
The mean-field energy and the coefficients are given by
 \begin{align}
  H_{MF}&= \frac{-JN}{2}\left(1 +\eta +2\delta +\lambda/4 \right),
  \\\label{solo1}  
  A_{\bold{k}}&=J\left[Q_{\bold{k}} + \lambda R_{\bold{k}}\right],
  \\ \label{solo2} 
 B_{\bold{k}}&= J\left[ S_{\bold{k}} + \lambda T_{\bold{k}}\right],
 \\\label{solo3}
 \varepsilon_{\bold{k}}&= \sqrt{A_{\bold{k}}^{2}-B_{\bold{k}}^{2}},\\\label{solo4} \nonumber 
 \end{align}
   where
 \begin{align}
 Q_{\bold{k}}&=\left[\left(1-\frac{\gamma_{k_x}}{2}\right) +\eta \left(1-\frac{\gamma_{k_y}}{2}\right)+\delta \left(2- \gamma_{k_x}\gamma_{k_y}\right) \right],\\
 \label{eqn3.21}
 S_{\bold{k}} &= \left[\frac{\gamma_{k_x}}{2} +\eta \frac{\gamma_{k_y}}{2}+\delta {\gamma_{k_x}\gamma_{k_y}}  \right], \\ \label{eqn3.22}
T_{\bold{k}}&= \frac{1}{4}(\gamma _{k_x} +\gamma_{k_y})- \frac{1}{4} \gamma _{k_x}\gamma_{k_y}, \quad  R_{\bold{k}}= \frac{1}{2}-T_{\bold{k}},
\\\label{eqn3.23}\nonumber
\end{align}
and the lattice structure constants are
\begin{equation}
\gamma_{k_x}=\cos k_{x}, \quad \gamma_{k_y}=   \cos k_{y}. 
\end{equation}
\section{Thermodynamic parameters}
The diagonalized quadratic Hamiltonian  \eqref{3d} gives the spin wave ground state and the excitation spectrum.  The energy spectrum is given by $\tilde{\varepsilon}_\bold{k} = 2\varepsilon_{\bold{k}}$. Figure \eqref{fig3.2} shows the plot of the spin wave energy long the direction $k_x=k_y$. In the top figure, the spectrum shows two zero modes at $\bold{k} = (0,0)$ and $\bold{k} = (\pi,\pi)$ for $\eta=1$, $\lambda = -2$ and several values of $\delta$. This result agrees with the result found previously, that no soft modes develop in the regime $\lambda>0$ (for the isotropic case $\eta=1$ and no NNN interaction) \cite{H}. Thus, the NNN interaction does not change the soft modes of the energy spectrum but only increases its peak. This is because the ring exchange term has already incorporated NNN sites. The bottom figure is interesting,  the zero (soft) mode of the energy at $\bold{k} = (\pi,\pi)$ is gapped as one moves away from $\eta=1$. 
\begin{figure}[h!]
\centering
\includegraphics[scale=0.25]{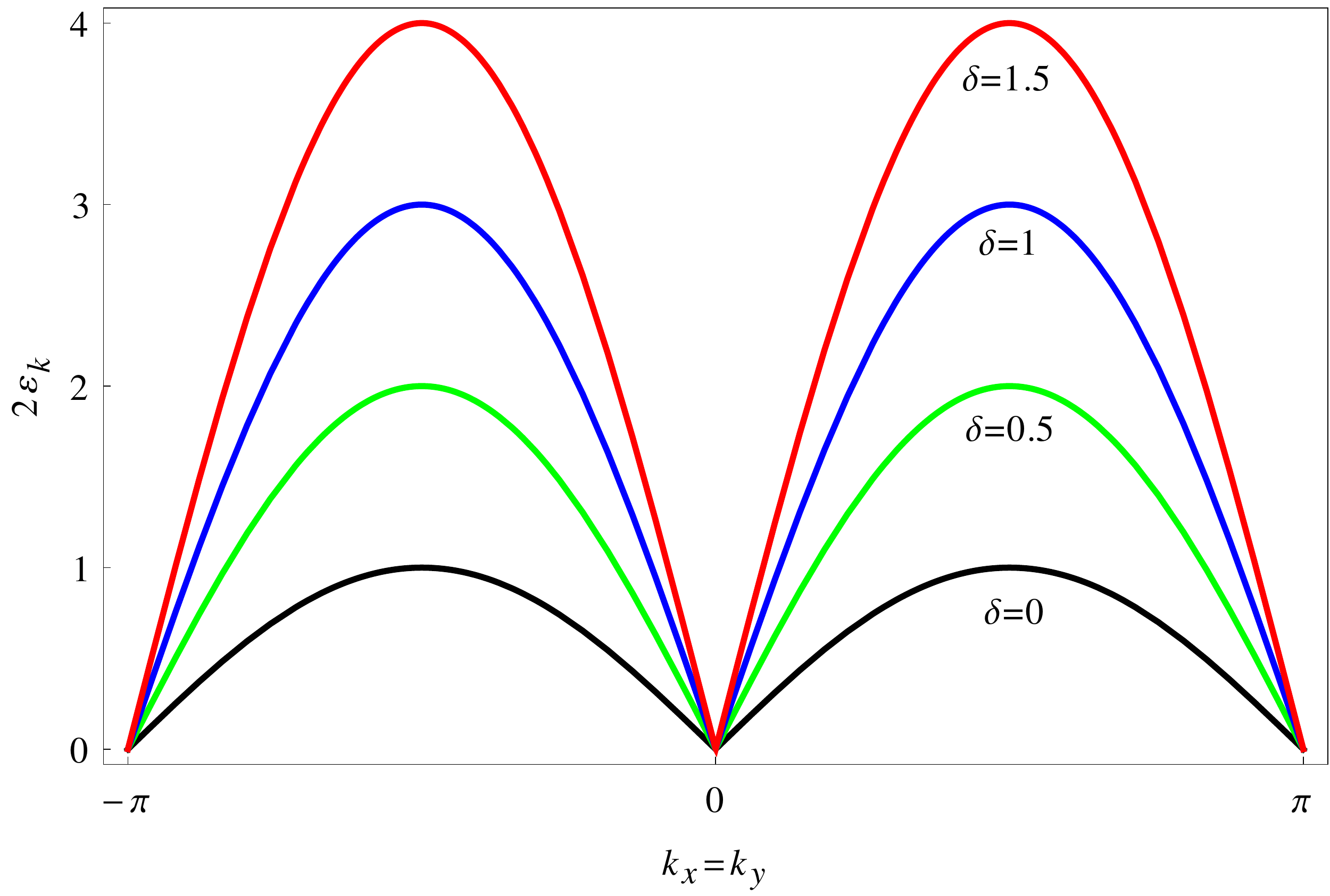}
\end{figure}
\begin{figure}[h!]
\centering
\includegraphics[scale=0.25]{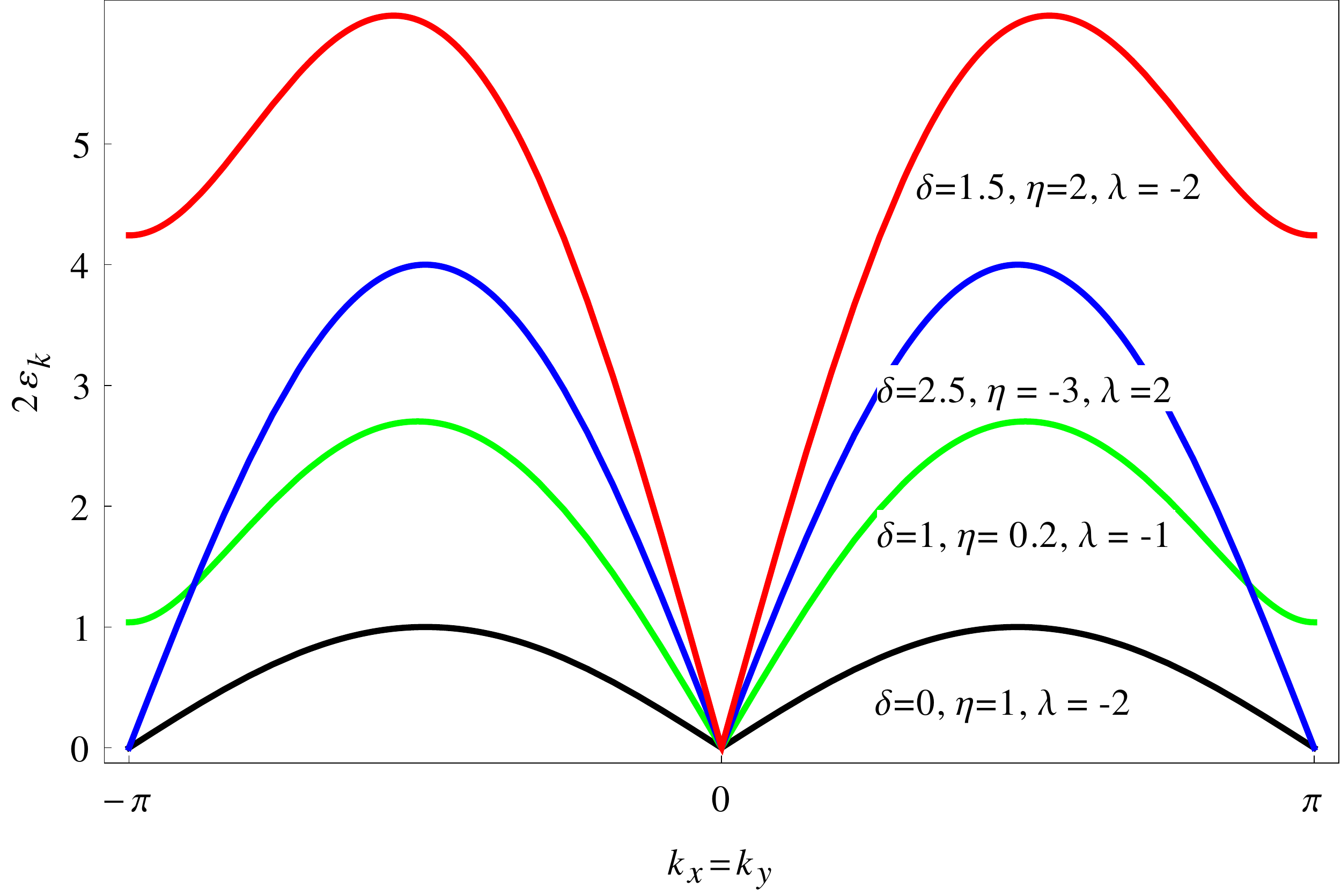}
\caption{(Color online): The energy spectrum along $k_x=k_y$ for $\eta=1$, $\lambda =-2$ and several values of $\delta$ (Top) and for several values of $\eta$, $\lambda$ and $\delta$ (Bottom). The destroyed soft mode at $\bold{k} = (\pi,\pi)$ for $\eta >1$ is restored whenever $\eta +\lambda +1 =0$.} \label{fig3.2}
\end{figure} 

However, we found that this soft mode is restored whenever $\eta +\lambda =-1$. To show how this condition came about, first consider the expansion of  the energy spectrum near the zero mode $\bold{k}=(0, 0)$:
\be
\tilde{\varepsilon}(\bold{k}\rightarrow 0)=\left[v_{x}^2 k_x^2 + v_{y}^2k_y^2\right]^{1/2}.
\label{s1}
\ee
The sound speeds $v_x$ and $v_y$ are expressed as
\begin{align}
v_x&= J\sqrt{(2+4\delta+ 2\eta +\lambda)(2\delta +1)}\\
v_y&= J\sqrt{(2+4\delta+ 2\eta +\lambda)(2\delta +\eta)}
\end{align}
Thus, the spectrum is gapless at $\bold{k}=0 $ independent of the values of  $\eta$, $\lambda$, and $\delta$ as shown in Fig.\eqref{fig3.2}. For a specific case $\eta=1$, $\delta=0$, and $\lambda =-2$, we have 
\be
\tilde{\varepsilon}(\bold{k}\rightarrow 0) = c\lvert\bold{k}\rvert,
\ee
where $c=J\sqrt{2}$.
Now consider the Taylor expansion of the energy spectrum near the zero mode $\bold{k}=(\pi, \pi)$:
\be
\begin{split}
\tilde{\varepsilon}(\bold{q}\rightarrow 0) & = \left[ v_{x}^2q_{x}^2 +  v_{y}^2q_{y}^2+ \Delta \tilde{\varepsilon} \right]^{1/2}
\end{split} 
\label{s2}
\ee
where $q_x = k_x-\pi$, etc. and the sound speeds $v_x$ and $v_y$ in this case are
\begin{align}
v_x&= \frac{J}{2}\sqrt{(2+4\delta+ 2\eta +\lambda)(2\delta -\lambda +1)}\\
v_y&= \frac{J}{2}\sqrt{(2+4\delta+ 2\eta +\lambda)(2\delta -\eta-\lambda)}
\end{align}
The energy gap $\Delta  \tilde{\varepsilon}$ is of the form
\be
\Delta \tilde{\varepsilon} =J^2(1+\eta +\lambda)(2+4\delta+ 2\eta +\lambda)
\ee
It's clear that at $\bold{q}=0$, no soft modes develop when $\eta +\lambda > 0$. This is illustrated in Fig.\eqref{fig3.2}(bottom). Notice that the spectrum  is imaginary in the regime $\eta +\lambda +1 <0 $, indicating a change in the ground state configuration.

We will now consider the thermodynamic quantities: Total internal energy and specific heat capacity of the spin waves.  At low temperature the spin wave are independent bosons and only low energy spin waves are excited. We can then use Eqs.\eqref{s1} and \eqref{s2} to determine the power of the temperature dependence on the specific heat $C_s$. In the regime where $\Delta\tilde{\varepsilon} =0$, i.e $1+\eta+\lambda = 0$. Eq. \eqref{s2} becomes
\be
\begin{split}
\tilde{\varepsilon}(\bold{q}) & = \left[ \varepsilon_{x}^2 +\varepsilon_{y}^2 \right]^{1/2} = \varepsilon,
\end{split} 
\label{s3}
\ee
where $\varepsilon_x = v_xq_x$ etc. The total energy of thermally excited spin waves is
\begin{align}
U &= \int\frac{ \tilde{\varepsilon}(\bold{q}) d \bold{q}}{e^{\tilde{\varepsilon}(\bold{q})/T}-1},\\
& = \frac{1}{(2\pi)^2v_xv_y}\int_{-\infty}^{\infty}d\varepsilon_x
\int_{-\infty}^{\infty}d\varepsilon_y \frac{\varepsilon}{e^{ \varepsilon /T}-1}\nonumber.
\end{align}
The above integrals can be converted into polar coordinate which gives
\begin{align}
U &= \frac{1}{2\pi v_xv_y}\int_{0}^{\infty}d\varepsilon
 \frac{\varepsilon ^2}{e^{ \varepsilon /T}-1}  = \zeta(3)\frac{T^3}{\pi v_xv_y}\nonumber .
\end{align}
The spin wave specific heat yields 
\be
C_s = \frac{\partial U}{\partial T}= \frac{3\zeta(3)}{\pi v_x v_y}T^2.
\ee 
%Another thermodynamic quantity of interest is the magnetization.  Physical observables are calculated by introducing the Green's function for the system. This is useful when one considers higher order expansions in  $1/S$ which involve bosonic interactions. 

The magnetization is easily derived by using the  zero temperature Green's function of the spin wave operator 
\be
G_{11}(\bold{k},t)= -i\langle T a_{\bold{k}}(t)a^{\dagger}_{\bold{k}}(0)\rangle,
\label{3bb}
\ee
where the subscript on $G$ refers to the first diagonal term in the total Green's function  matrix \cite{Z3}. Using  \eqref{3b}, the Fourier transform of \eqref{3bb} gives
\be
G_{11}(\bold{k},\omega)= \frac{\omega + A_{\bold{k}} -i\eta}{(\omega -\tilde{\varepsilon}_{\bold{k}} + i\eta)(\omega +\tilde{\varepsilon}_{\bold{k}} - i\eta)}  ,
\label{3bc}
\ee
The average magnetization is then expressed as
\begin{align}
M &= \frac{1}{2}-\langle a^{\dagger}_ia_i\rangle \\
&= \frac{1}{2}+\frac{1}{N}\sum_{\bold{k}}\int_{-\infty}^{\infty} \frac{d\omega}{2\pi i} G_{11}(\bold{k}, \omega)e^{-i\omega 0^{-}}\nonumber.
\end{align}
\begin{figure}[h!]
\centering
\includegraphics[scale=0.25]{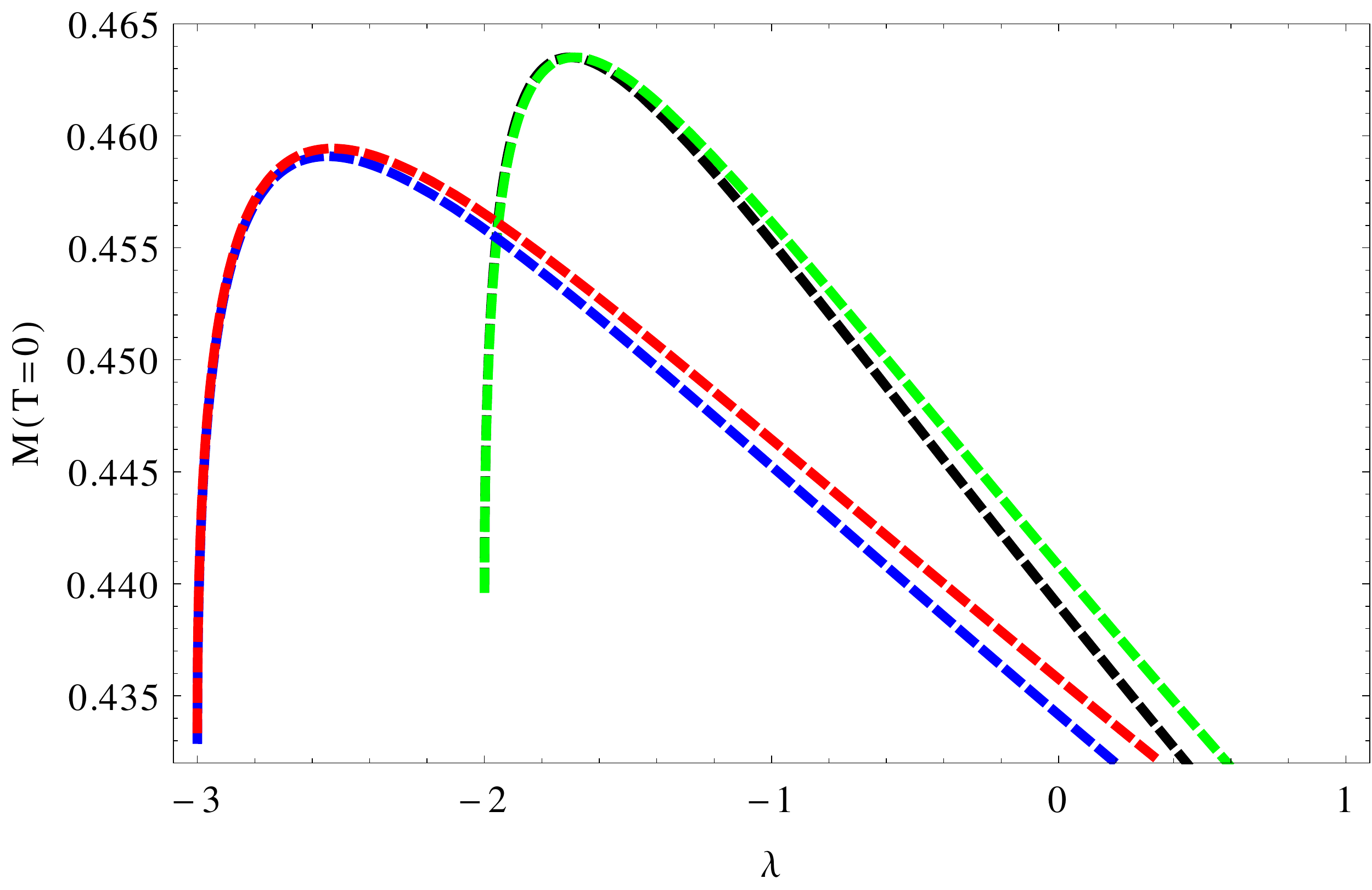}
\caption{(Color online): Magnetization vs $\lambda$ for the isotropic case $\eta=1$ with $\delta=0 (\text{black})$, $\delta=0.029(\text{green})$  and the anisotropic case $\eta=2$ with $\delta=0 (\text{blue})$, $\delta=0.029(\text{red})$. The value of the magnetization for $\eta=1$, $\delta = 0$ and $\lambda = 0$ is $M=0.439$ and $M=0.434$ for $\eta=2$, $\delta$ and $\lambda$ remains the same. The decrease is $1.1\%$ of $M=0.439$. The magnetization drops rapidly when $\eta +\lambda +1 =0$.} \label{fig3.3}
\end{figure} 
Evaluating the second term by contour integration, we obtain the deviation
\be
 \Delta M = \frac{1}{2N}\sum_{\bold{k}}\left[\frac{A_{\bold{k}}}{\sqrt{A_{\bold{k}}^2-B_{\bold{k}}^2}}-1\right],
\ee
where $A_{\bold{k}}$ and $B_{\bold{k}}$ are given by Eq.\eqref{solo1} and Eq.\eqref{solo2} respectively. In Fig. \eqref{fig3.3}, the magnetization is plotted against $\lambda$ for the isotropic $\eta=1$ and the anisotropic cases $\eta=2$  with two different values of $\delta$. The magnetization decreases rapidly  when $\eta + \lambda = -1$ reflecting the fact that no soft modes is developed in the regime $\eta + \lambda >0 $. For $\delta = 0$ and  $\lambda = 0$, anisotropy decreases the value of the isotropic magnetization  $M=0.439$\cite{K, Z2} by $1.1\%$.

% The magnetization drops rapidly as it approaches $\lambda=-2$ in agreement with the constraint $\eta +\lambda +1 =0$. However, the maximum value of the magnetization decreases as $\delta$ increases.
%\begin{figure}[h!]
%\centering
%\includegraphics[scale=0.25]{mag_1.pdf}
%\end{figure}

\section{Spin stiffness}
The spin stiffness, or helicity modulus is the change in the ground state of a spin system resulting from applying a twist $\phi$ between every pair of neighbouring lattice bonds. It is given by the second derivative of the twisted ground state energy with respect to the twist
\be
\rho_s =\left.\frac{1}{N}\frac{d^2 E_0({\phi})}{d\phi^2}\right|_{\phi=0},
\label{3e}
\ee

where $E_0(\phi)$ is the ground state energy of the twisted Hamiltonian and $N$ is the number of lattice sites of the system. In the thermodynamic limit, the sign of the spin stiffness determines if the system has a long-range magnetic order(LRMO), $\rho >0$ means the existence of LRMO in the system whereas $\rho_s=0$ means no LRMO\cite{Y}. The above mathematical expression for $\rho_s$ is equivalent to the difference in the ground state energies
\be
\frac{E_0(\phi)-E_0}{N}= \frac{1}{2}\rho_s\phi^2,
\ee
where $E_0(\phi)\equiv  \langle H(\phi)\rangle$ and $E_0\equiv  \langle H_0\rangle$. The twist dependent Hamiltonian is found by rotating the spin at site $i$ by $\phi_i$ around the $z$-axis i.e $S_i^{+}\rightarrow S_{i}^{+}e^{i\phi_i}$, $S_i^{-}\rightarrow S_{i}^{-}e^{-i\phi_i}$, so $H(\phi)$ is given by 
\begin{align}
H(\phi) =&-J\sum_{\left\langle ij \right\rangle} \left(S^{+}_{i}S^{-}_{j}e^{i(\phi_i-\phi_j)} +h.c \right)\nonumber \\&-J^{\prime}\sum_{\left\langle jk \right\rangle} \left(S^{+}_{j}S^{-}_{k}e^{i(\phi_j-\phi_k)} +h.c \right)\nonumber\\
&-J_D\sum_{\left[ik \right]}\left(S^{+}_{i}S^{-}_{k}e^{i(\phi_i-\phi_k)} +h.c \right)\label{3f} 
\\&-K\sum_{\left\langle ijkl  \right\rangle} \left(S_{i}^{+}S_{j}^{-}S_{k}^{+}S_{l}^{-}e^{i(\phi_i-\phi_j+\phi_k-\phi_l)} + h.c\right)\nonumber.
\end{align}  
Now consider a uniform twist $\phi_x$, along the positive $x$-axis nearest neighbour bonds. Using the labelling in fig.\eqref{fig3} we have $\phi_i-\phi_j=\phi_x =\phi_l-\phi_k$, $\phi_j-\phi_k=0= \phi_i-\phi_l$, and $\phi_i-\phi_k=\phi_x$. Thus, the twist dependence on the second and the fourth terms in \eqref{3f} vanishes. MacLaurin expansion of \eqref{3f} around $\phi_x=0$ gives
\be
H(\phi_x)= H_0 + \phi_x j_x^s-\frac{1}{2}\phi_x^2 T_x^s.
\ee
The spin current operator $j_x^s$ (paramagnetic term) and the spin kinetic energy operator $T_x^s$ (diamagnetic term)  are defined mathematically as
\be
j_x^s =\left.\frac{d H({\phi_x})}{d\phi_x}\right|_{\phi_x=0} , \quad{ \left.T_x^s=- \frac{d^2 H({\phi_x})}{d\phi_x^2}\right|_{\phi_x=0}}.
\ee 
Explicitly we have
\be
j_x^s=i\sum_{l}\left[-JS_l^{+}S_{l+\hat{x}}^{-}-J_D\left(S_l^{+}S_{l+\hat{x}+\hat{y}}^{-} +S_l^{+}S_{l+\hat{x}-\hat{y}}^{-}\right)- h.c\right]
\label{3h}
\ee
\be
T_x^s=\sum_{l}\left[-JS_l^{+}S_{l+\hat{x}}^{-}-J_D\left(S_l^{+}S_{l+\hat{x}+\hat{y}}^{-} +S_l^{+}S_{l+\hat{x}-\hat{y}}^{-}\right) +h.c\right].
\label{3i}
\ee

Using second order perturbation theory we obtain \cite{Y,Z}
\be
\rho_s= \frac{2}{N}\left[\frac{1}{2}\langle-T_x^s\rangle -\sum_{\nu\neq0}\frac{\lvert\braket{0|j_x^s|\nu}\rvert^2}{E_{\nu}-E_0}\right],
\label{3g}
\ee
where $\ket{0}$ and $\ket{\nu}$ are the ground state and the excited states respectively. 

\begin{figure}[h!]
\centering
\includegraphics[scale=0.25]{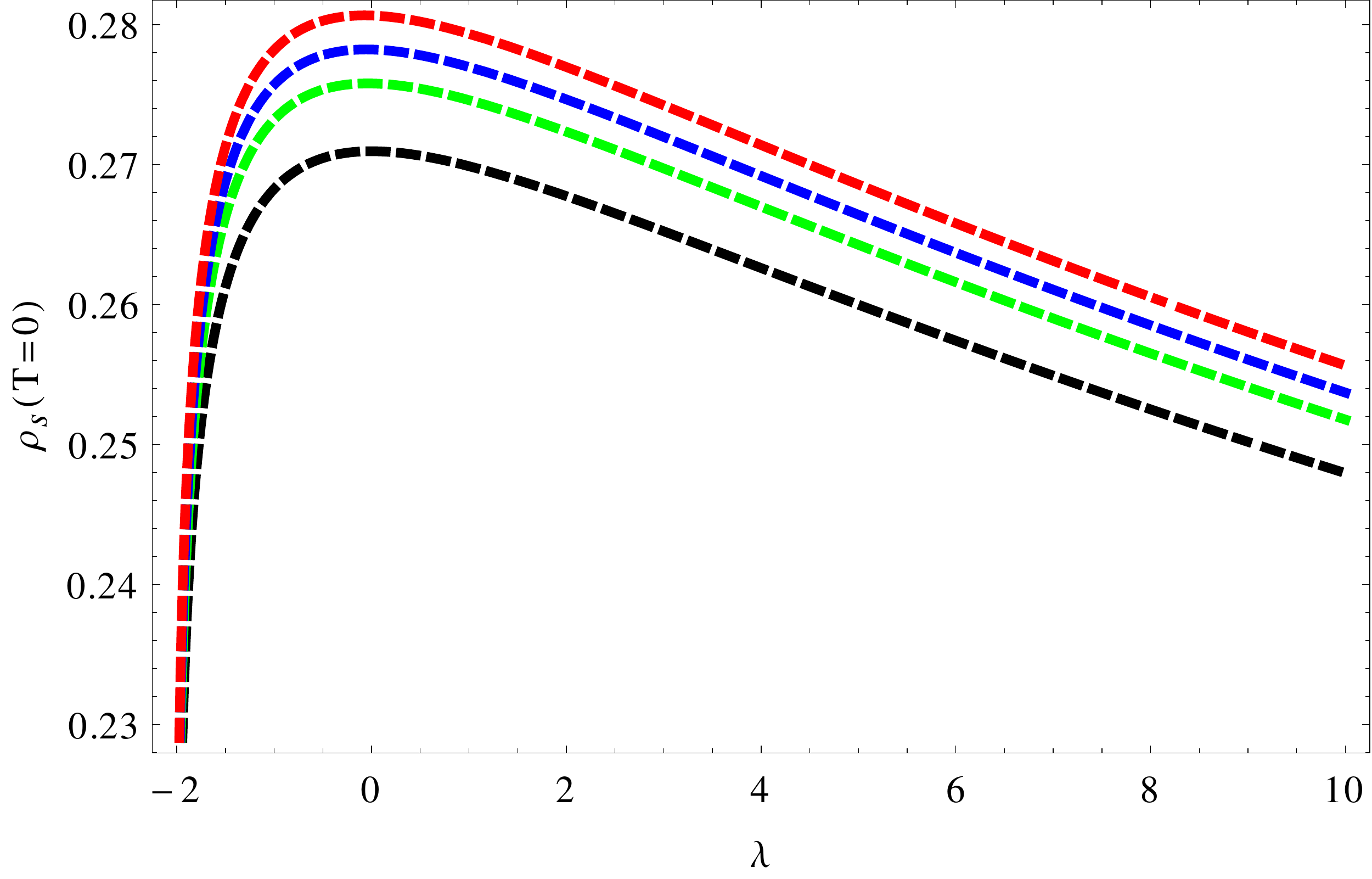}
\end{figure}
\begin{figure}[h!]
\centering
\includegraphics[scale=0.25]{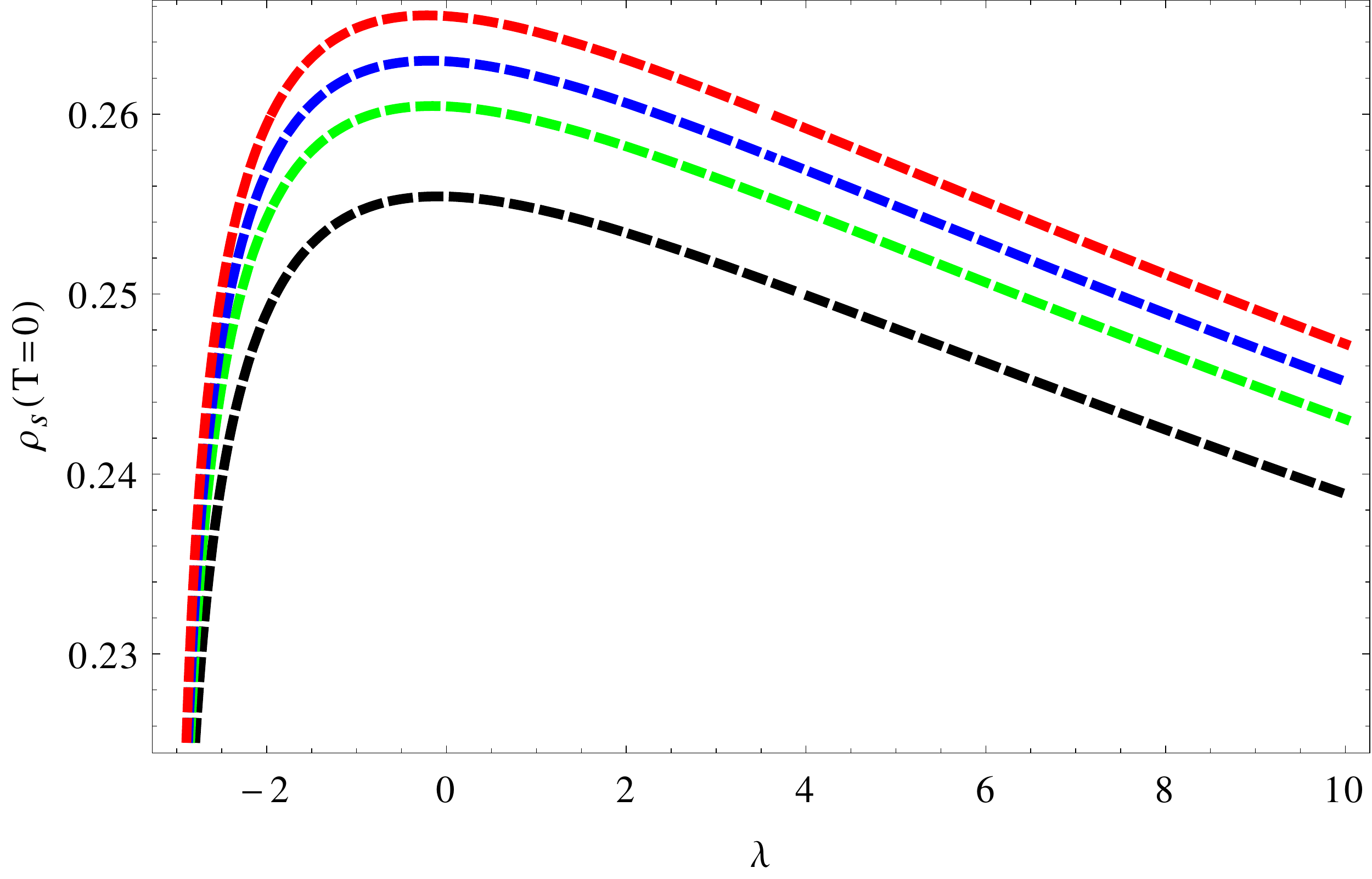}
\caption{(Color online): Spin stiffness vs $\lambda$ for isotopic case  (top) $\eta =1$, $\delta = 0 (\text{black})$, $0.01(\text{green})$, $0.015(\text{blue})$, $0.02(\text{red})$. Anisotropic case (bottom) $\eta =2$, $\delta$ the same. Anisotropy decreases the value of the spin stiffness. The stiffness shows a rapid decrease when the constraint $\eta +\lambda +1 =0$ is satisfied.} \label{fig4}
\end{figure}
At zero temperature the second term in \eqref{3g} vanishes. The first term can be written in terms of the bosonic representations \eqref{3} and the Bogoluibov transformations \eqref{3b} which gives
\be
\rho_s(T=0) =\frac{J}{2}(1+2\delta) +\frac{1}{N}\sum_{\bold{k}}\left(C_{\bold{k}} -\frac{A_{\bold{k}}C_{\bold{k}}-B_{\bold{k}}D_{\bold{k}}}{\varepsilon_{\bold{k}}}\right),
\label{3j}
\ee
where the new coefficients are
\be
C_{\bold{k}}= JQ_{\bold{K}}(\eta=0), \quad D_{\bold{k}}= JS_{\bold{K}}(\eta=0).
\ee
Figure \eqref{fig4} shows the plot of the spin stiffness against the parameter of the ring exchange $\lambda$ for two cases: Isotropic case (top) $\eta=1$ and anisotropic case $\eta=2$(bottom) for different values of the NNN exchange $\delta$. In the isotropic case we recover exactly the result obtained in Ref.[10] for $\eta=1$ and $\delta =0$. As $\delta$ increases above zero, the value of the stiffness increases with its maximum centers at $\lambda =0$. This is obvious from the first term in Eq.\eqref{3j}. The anisotropy case $\eta=2$ is quite interesting. The spin stiffness decreases by $5.7 \%$ of its isotropic maximum value independent of the value of $\delta$. The maximum values of $\rho_s$ still center at $\lambda=0$ in this case.  Notice that the spin stiffness shows a sharp decrease when $\eta +\lambda +1 =0$ which indicates an ordered phase with $(\pi,\pi)$ symmetry. The nature of this ordered phase for the case $\eta=1$ and $\lambda= -2$  has been suggested to be a supersolid phase \cite{H}. In general this phase transition will occur in the regime where $\eta + \lambda +1 <0$.

\begin{figure}[h!]
\centering
\includegraphics[scale=0.25]{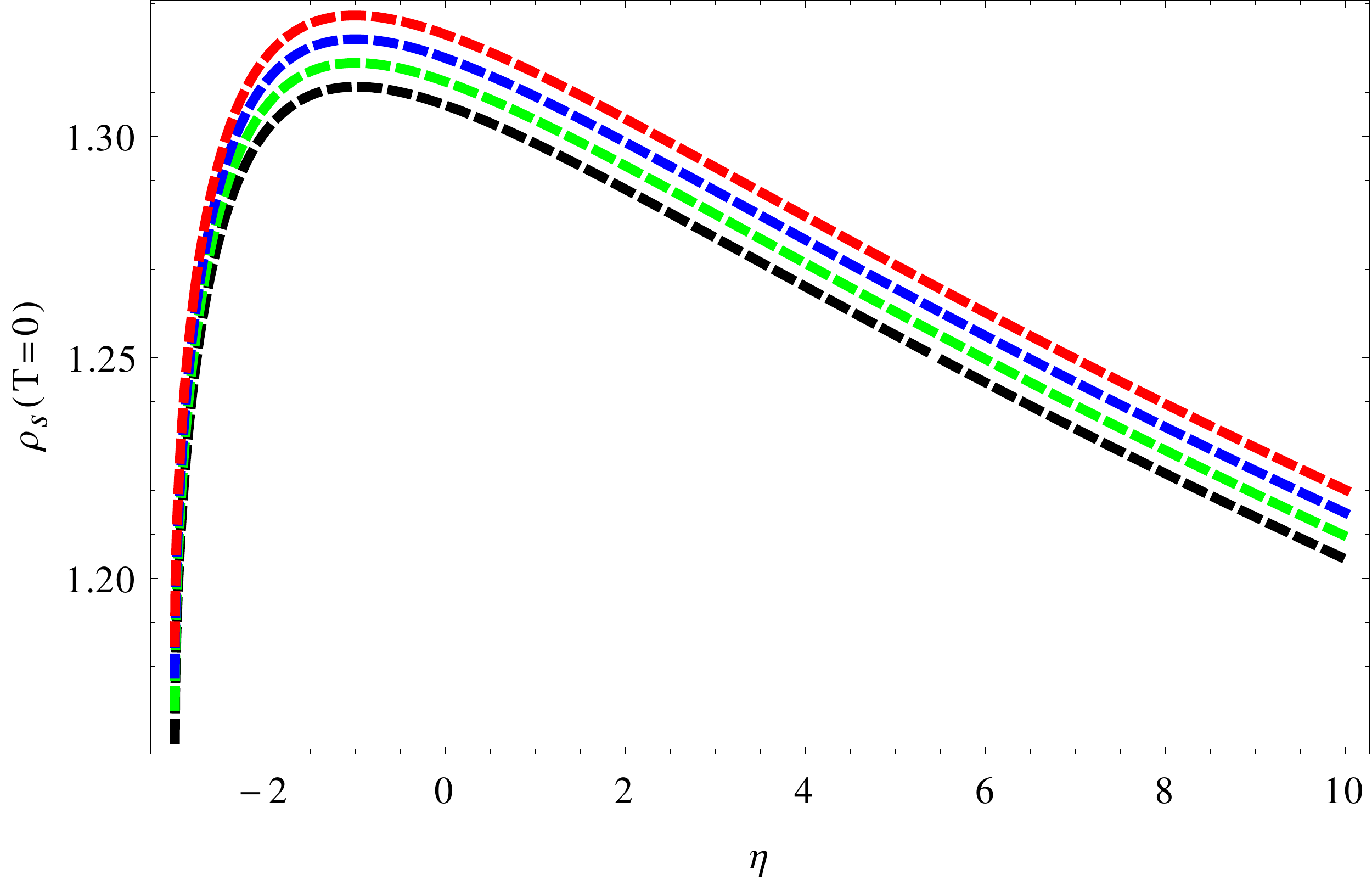}
\caption{(Color online): Spin stiffness vs $\eta$ for $\lambda =2$, and $\delta = 2(\text{black})$, $2.01(\text{green})$, $2.02(\text{blue})$, $2.03(\text{red})$. The stiffness shows a rapid decrease when the constraint $\eta +\lambda +1 =0$ is satisfied.} \label{fig5}
\end{figure}

On the other hand, the plot of spin stiffness against $\eta$ shows a similar trend. It has a maximum at $\eta=0$ and decreases as $\eta$ moves away from zero for some values of $\delta$. It also decreases so fast as it approaches the region where $\eta +\lambda +1 =0$ (see Fig.\eqref{fig5}).
\section{conclusion}
In this paper we have studied the hard-core boson ( zero field $XY$ model) using linear spin wave theory. In contrast to the previous study of this model, we included the effects of NN, NNN anisotropy and ring exchange interactions which introduced three dimensionless parameter as opposed to just one. It was shown that no soft modes (Goldstone modes) develop in the regime $\eta + \lambda > 0$.  The spin stiffness was obtained by applying a twist along the $x$-axis nearest neighbour sites. We showed that anisotropy decreases the values of the stiffness by $5.7\%$ of its isotropic maximum value for specific values of $\lambda$ and $\delta$. Similar reduction was calculated for the magnetization.  The stiffness shows  a sharp decreases as it approaches $\eta + \lambda =-1$ reflecting the fact that no soft modes develops for $\eta + \lambda > 0$. In general the supersolid phase with $(\pi, \pi)$ symmetry suggested in the previous work on this model will occur in the regime $\eta + \lambda +1 < 0$.
%\section*{Appendix: Spin waves specific heat at low temperature}

\end{document}